\documentclass[final,authoryear,3p,times,twocolumn]{elsarticle}

\usepackage{graphics}
\usepackage{epsfig}

\usepackage{amssymb}

\journal{New Astronomy Reviews}

\begin{document}

\begin{frontmatter}



\title{What broad emission lines tell us about how active galactic nuclei work}


\author{C. Martin Gaskell}

\address{Astronomy Department, University of Texas, Austin, TX 78712, USA}

\begin{abstract}
I review progress made in understanding the nature of the broad-line region (BLR) of active galactic nuclei (AGNs) and the role BLRs play in the AGN phenomenon.  The high equivalent widths of the lines imply a high BLR covering factor, and the absence of clear evidence for absorption by the BLR means that the BLR has a flattened distribution and that we always view it near pole-on.  The BLR gas is strongly self-shielding near the equatorial plane.  Velocity-resolved reverberation mapping has long strongly excluded significant outflow of the BLR and shows instead that the predominant motions are Keplerian with large turbulence and a significant net inflow.  The rotation and turbulence are consistent with the inferred geometry. The blueshifting of high-ionization lines is a consequence of scattering off inflowing material rather than the result of an outflowing wind.  The rate of inflow of the BLR is sufficient to provide the accretion rate needed to power the AGN.  Because the motions of the BLR are gravitationally dominated, and the BLR structure is very similar in most AGNs, consistent black hole masses can be determined. The good correlation between these estimates and masses predicted from the bulge luminosities of host galaxies provides strong support for the similarity of AGN continuum shapes and the correctness of the BLR picture presented.  It is concluded that although many mysteries remain about the details of how AGNs work, a general overall picture of the torus and BLR is becoming clear.

\end{abstract}


\end{frontmatter}


\section{Introduction}
\label{}

A broad-line region (BLR) is present in all AGNs accreting at moderate- to high-Eddington ratios.  BLRs are important both because they are our best probe of how AGNs work and because of their potential for readily providing masses of supermassive black holes (SMBHs) back to the earliest times of galaxy formation.  However, in order to be able to use BLRs to reliably estimate the masses of SMBHs it is essential to understand the structure and kinematics of BLRs.   Over the last four decades there have been wide-ranging and, not infrequently, mutually contradictory views of the nature of the BLR (see reviews by \citealt{mathews+capriotti85}, \citealt{osterbrock+mathews86}, and \citealt{sulentic+00}).  However, I believe that the situation is improving.  I review here what I consider to be the clearest pointers to the underlying structure and kinematics of the BLR and I argue that, while there are certainly many interesting problems remaining, the basic picture is now becoming fairly secure.  I furthermore believe that this picture applies to {\em all} BLRs because BLR equivalent widths and line ratios are remarkably similar, especially in the ultraviolet.

\section{The structure of the broad-line region and torus}

The two most basic questions about the BLR are ``what does it look like?'' and ``how is it moving?''  The traditional picture of the BLR of an AGN for over 40 years (and one which is widely depicted in cartoons of AGNs) has been that there is a central source emitting ionizing radiation roughly spherically, and that it is surrounded by a roughly spherical mist of cloudlets.  This is depicted in the left-hand-panel of Fig.\@ 1.  Each individual cloud, if it is big enough, will have a structure as shown in the right-hand-panel of Fig.\@ 1.  It will be highly ionized on the front, and if it has a high-enough column density, it will be mostly neutral on the back.  The front emits high-ionization lines such as He\,II, He\,I, O\,VI, N\,V, and C\,IV, while the back emits low-ionization lines such as Mg\,II, Ca\,II, O\,I, and Fe\,II.  All these lines are well-known in AGNs.

\begin{figure}[t!]
\resizebox{\hsize}{!}{\includegraphics{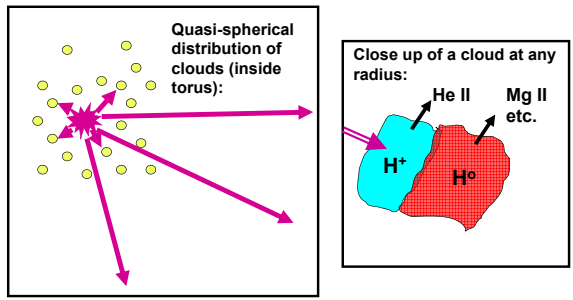}}
\caption{ \footnotesize The left frame shows a cartoon of a common traditional view of the BLR.  The right frame shows a schematic close-up of an individual cloudlet.} \label{}
\end{figure}

The emissivity of each line as a function of distance from the front of the cloud can be calculated with the photoionization code CLOUDY \citep{ferland+98}.  Fig.\@ 2 shows emissivities for some well-known lines.

\begin{figure}[]
\resizebox{\hsize}{!}{\includegraphics{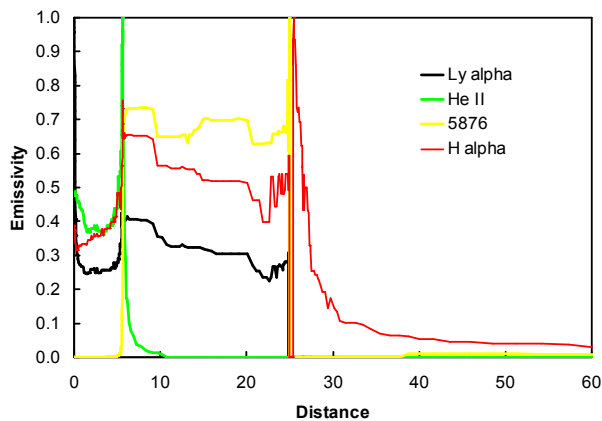}}
\caption{ \footnotesize Relative emissivity of four lines in a typical BLR cloudlet (normalized to a maximum emissivity of one for each line) versus distance (in arbitrary units) from the ionized face of the cloud..} \label{}
\end{figure}

\citet{Baldwin+95} showed that the sum of contributions from clouds with a distribution of cloud properties (densities and distances from the center) will automatically produce a total spectrum similar to what is observed from AGNs.   This is the so-called LOC model.\footnote{Ostensibly from ``locally optimally emitting clouds.''}  This was important because it showed that no ``fine-tuning'' of cloud conditions was needed to explain AGN spectra.

Despite the success of the traditional picture in general, and the LOC model in particular, in explaining the overall spectrum of an AGN, the problem with this picture (see \citealt{gaskell+07}) is that to explain the strengths of the BLR lines the covering factor has to be large (50\% or so), yet if this is so, and if the cloudlets are covering the central source uniformly, we ought to see Lyman continuum absorption by the BLR clouds.  In fact Lyman continuum absorption due to the BLR is {\em never} convincingly seen (\citealt{antonucci+89} - see discussion in \citealt{macalpine03} and \citealt{gaskell+07}).  We believe, as proposed by S. Phinney (see \citealt{antonucci+89}), that the need for a high covering factor plus the lack of Lyman continuum absorption {\em requires the BLR to have a flattened distribution and requires us to be viewing it through a hole.}  This conclusion is supported by recovery of what is called the ``transfer function'' of some lines (the transfer function is the temporal response of a line to a delta-function event in continuum light curve).  Transfer functions for low-ionization lines have always implied that there is little or no gas along the line of sight \citep{krolik+91,horne+91,mannucci+92,pijpers+wanders94}, and thus that at least the low-ionization gas in the BLR has a flattened distribution.

Having a high overall covering factor but a flattened distribution means that near the equatorial plane there will be a close to a 100\% chance that any path will intersect a BLR cloud.  The clouds will thus be self-shielding.  Radiation from the central source can freely escape near the axis of symmetry, but is strongly diluted in the equatorial plane.  This is schematically illustrated in Fig.\@ 3.

\begin{figure}[]
\resizebox{\hsize}{!}{\includegraphics{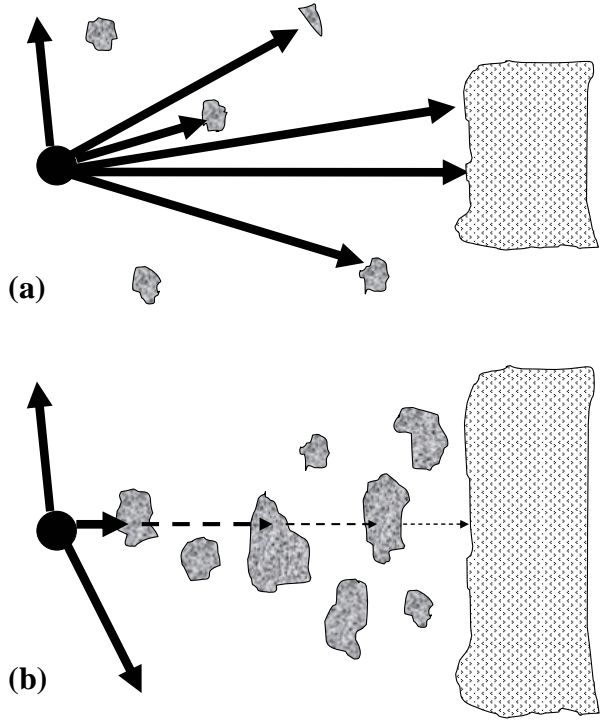}}
\caption{ \footnotesize Schematic cross section of the BLR and torus in a plane through the axis of symmetry.  The torus is on the right.  Ionizing radiation is attenuated in the equatorial plane, but can freely escape near the poles. Figure from \citet{gaskell+07}} \label{}
\end{figure}

It is easy to calculate the average radial dependence of the ionization and the emissivities of all the lines coming from cloudlets with a distribution such as in Fig.\@ 3.  The ionization structure of a single cloud in CLOUDY is now spread out in radius as illustrated schematically in Fig.\@ 4.  The horizontal axis in Fig.\@ 2 can now be read as distance into the BLR rather than distance into an individual cloud.  Our model is in fact very similar to the old ``filling factor'' model of \citet{macalpine72}.

\begin{figure}[]
\resizebox{\hsize}{!}{\includegraphics{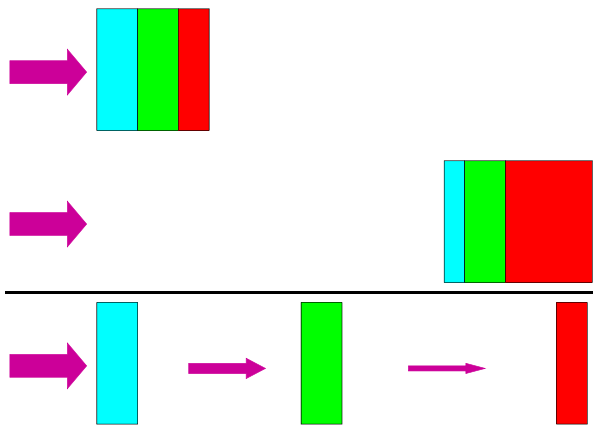}}
\caption{ \footnotesize Cartoon of the relationship between a traditional cloudlet model (top two thirds of the diagram) and the self-shielding model (bottom third).  The different shadings symbolize three regions producing lines of differing degrees of ionization (cf. Fig.\@ 3). .} \label{}
\end{figure}

The earliest reverberation mapping of multiple lines \citep{gaskell+sparke86} showed that the high-ionization lines were coming from smaller radii than the low-ionization lines.  High-ionization lines were also wider (e.g., \citealt{shuder82,mathews+wampler85}).  The radial ionization stratification of the BLR has been well confirmed by later reverberation mapping.  The best reverberation-mapped AGN is NGC~5548.
The horizontal axis of Fig.\@ 5 (taken from \citealt{gaskell_goosmann_klimek08}) shows the reverberation-mapping time lags (i.e., the effective radii) for lines of a variety of ions from \citet{clavel+91}, \citet{peterson+91}, and \citet{bottorff+02}. (See table in \citealt{gaskell+07} for details).  The observed lags cover an order of magnitude in radius.  NGC\,5548 is not unique in this regard: an identical range of radii has also been found for Mrk~110 by \citet{kollatschny03}.  The vertical axis shows the lags predicted for the same lines by the LOC model \citep{korista+goad00,bottorff+02}.  It can be seen that while there is a correlation, these predicted lags cover a much smaller range of radii.  The reason for this can be appreciated in Fig.\@ 4.  In the LOC model (top part of the figure) every cloud has a highly-ionized front part; the clouds just differ in the degree of ionization.

The self-shielding model \citep{gaskell+07} solves the problem of why the ionization stratification is so strong.  In the self-shielded model (bottom of Fig.\@ 4) there is a clear spatial separation of the differing ionizations.  It can be seen in Fig.\@ 5 that the self-shielding model gives good agreement with the observed lags.

\begin{figure}[]
\resizebox{\hsize}{!}{\includegraphics{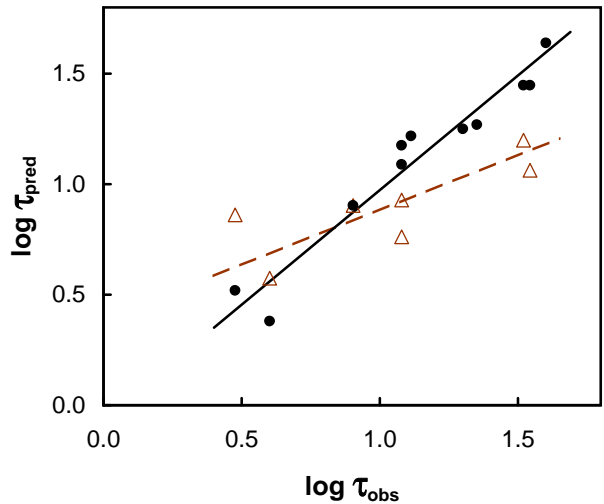}}
\caption{ \footnotesize Observed lags ($\tau_{obs}$) versus predicted lags ($\tau_{pred}$) for NGC~5548. The open triangles are the predictions of the LOC model, and the solid dots are the predictions of the self-shielding model of \citet{gaskell+07}.  The lines are OLS-bisector fits \citep{isobe+90}. Figure from \citet{gaskell_goosmann_klimek08}.}
\label{small_cartoon}
\end{figure}

\citet{netzer+laor93} made the important suggestion that the outer edge of the BLR coincided with the dust sublimation radius of the torus.  Reverberation mapping observations show that the low-ionization gas in the BLR indeed extends out to the dusty torus \citep{suganuma+06,gaskell+07}.  The covering factor of the torus can be calculated statistically from the ratio of type-1 (face-on) to type-2 (edge-on) AGNs, and directly for individual objects from the strength of the thermal emission (e.g., \citealt{maiolino+07}).  We argue \citep{gaskell+07} that the covering factors of the BLR and torus have to be the same.  This is because if the {\em torus} has a lower covering factor than the BLR we would see the BLR in absorption against the central continuum source in some objects near the type-2 viewing position.  This is never seen.  On the other hand, if the {\em BLR} has a lower covering factor, some of the dusty torus will see direct radiation from the central source.  This cannot be the case for much of the torus because it would then be unable to exist as close in as is seen.

The overall picture we get of the torus and BLR is indicated schematically in the cartoon in Fig.\@ 6 and in the computer generated renditions shown in Fig.\@ 7.  The best description of the appearance is to say that the BLR and torus look like a bird's nest.  This picture is identical to that favored for totally independent reasons in an unfortunately almost totally overlooked paper by \citet{mannucci+92}.  They inferred a ``bird's nest'' geometry from a combined analysis of line profiles and transfer functions in NGC 5548.  The positions of masers in NGC 1068 also provides support for a thick BLR-torus \citep{greenhill+96}.

\begin{figure}[]
\resizebox{\hsize}{!}{\includegraphics{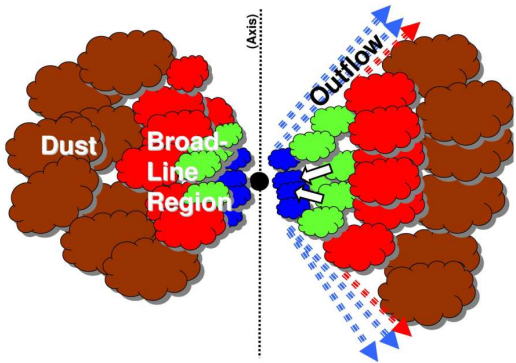}}
\caption{ \footnotesize A schematic view of the BLR and torus of an AGN in a plane through the axis.  The figure is approximately to scale (except that the black hole is shown too large.) Figure from \citet{gaskell+07}.} \label{small_cartoon}
\end{figure}

\begin{figure}[]
\resizebox{\hsize}{!}{\includegraphics{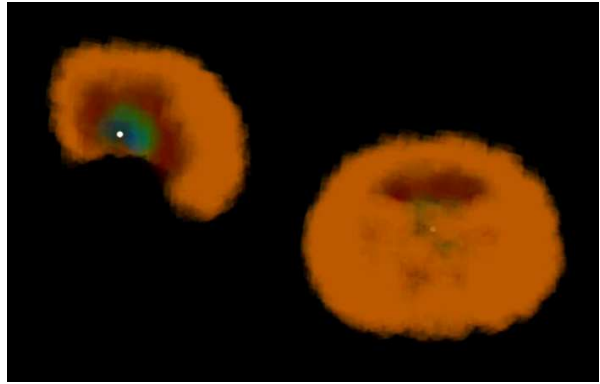}}
\caption{ \footnotesize Computer-generated renditions of the appearance of the BLR and clumpy torus of an AGN seen from a ``Seyfert 1.9'' viewing position (lower right), and a cutaway view from the same position (upper left).  The cut makes a 45 degree angle to the projection of the line of sight onto the equatorial plane. Computer renditions by Daniel Gaskell.} \label{}
\end{figure}

\section{THE KINEMATICS OF THE BLR AND TORUS}

\subsection{Determining the Direction of Motion}

The kinematics of the BLR have been a long-standing problem.  It has been known from the earliest days of AGN studies that the lines are very broad (for a review of the earliest literature see \citealt{seyfert43}), but Doppler shifts only tell us the motion of gas along the line of sight.  To know whether the gas is inflowing, outflowing, moving in random virialized orbits, or in more planar Keplerian orbits in a disc we need to know the line-of-sight velocity as a function of position relative to the black hole.

The discovery of narrow intrinsic absorption in NGC 4151 \citep{mayall34,anderson+kraft69} and of broad absorption lines (BALs) in PHL 5200 \citep{lynds67} proved that {\em some} gas was outflowing from AGNs.  However, BALs commonly extend to velocities several times higher than those observed for the BLR in the same objects (see, for example, \citealt{turnshek+88}), so it is not clear that there is necessarily any connection between BALs and BLRs.  The case for an outflowing BLR was strengthened though when \citet{blumenthal+mathews75} and \citet{baldwin75} showed that a radiatively-accelerated outflow could reproduce the observed line profiles well in some objects.  However, \citet{capriotti+80} showed that other models could provide comparably good fits to broad-line profiles, and so demonstrated that fits to individual line profiles alone could not uniquely determine the kinematics.

More progress was made by comparing lines of differing ionizations.  \citet{gaskell82} discovered that high-ionization broad lines were blueshifted with respect to low-ionization lines, and pointed out that this requires there to be {\em radial} motions plus some source of opacity.  This blueshifting has now been widely confirmed.  \citet{gaskell82} suggested that the blueshifting could be the result of a ``disk-wind'' model where the high-ionization lines arise in a wind outflowing above the accretion disc.  \citet{wilkes+carswell82} pointed out a problem with any purely radial motion: the profiles of C IV and Lyman $\alpha$ were observed to be very similar, yet, for optically-thick clouds, Lyman $\alpha$ is emitted very anisotropically.  To satisfy this constraint the clouds either had to be optically thin, or not moving purely radially.

Obviously the question of the direction of motion could be settled if it could be determined which gas was on which side of the black holes.  The best way of doing this is through velocity-resolved reverberation mapping \citep{gaskell88}.  How this works is illustrated in Fig.\@ 8.  Surprisingly, velocity-resolved reverberation mapping results \citep{gaskell88,koratkar+gaskell89,crenshaw+blackwell90,koratkar+gaskell91a,koratkar+gaskell91b, koratkar+gaskell91c,korista+95,done+krolik96,ulrich+horne96,sergeev+99} strongly ruled out significant outflow of both high- and low-ionization lines (see example in Fig.\@ 9).

\begin{figure}[]
\resizebox{\hsize}{!}{\includegraphics{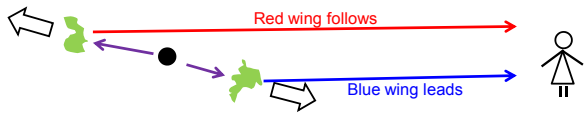}}
\caption{ \footnotesize Velocity-resolved reverberation mapping.  Because of light-travel-time effects, the gas on the near side of the AGN is seen to respond to continuum changes first.  For the the hypothetical outflowing BLR illustrated here, the {\em blue} wing of a line would vary first.} \label{}
\end{figure}

\begin{figure}[]
\resizebox{\hsize}{!}{\includegraphics{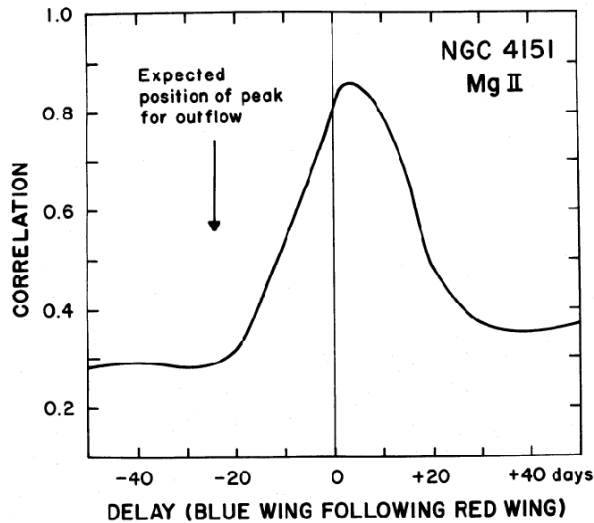}}
\caption{ \footnotesize The cross-correlation function for the blue and red wings of the Mg II line in NGC~4151 as a function of time delay.  The predicted peak in the correlation function for pure outflow (blue wing varies first) is shown by the arrow.  It can be seen instead that the strongest correlation is for near zero delay (what is expected for virialized or Keplerian motion), but with the red wind leading by a small but significant amount thus implying some net inflow.  Figure from \citet{gaskell88}.} \label{}
\end{figure}

Ruling out significant outflow of the BLR was important not just because of what it said about how AGNs work, but also because it meant that the BLR motions were {\em gravitationally dominated}.  The BLR could thus be used for determining the masses of the central black holes.  This permitted the first reverberation mapping determinations of black hole masses and Eddington ratios \citep{gaskell88,koratkar+gaskell89,crenshaw+blackwell90,koratkar+gaskell91a, koratkar+gaskell91b, koratkar+gaskell91c}.\footnote{The BLR was first used to estimate masses of AGN black holes by \citet{dibai77} who estimated BLR sizes using photoionization considerations.  At that stage, of course, there was no clear evidence that the BLR was virialized.  See \citet{bochkarev+gaskell09}.}  Note that while gravity dominates BLR motions, the simple fact that radiation pressure was at one time considered to be driving BLR motions (e.g., \citealt{blumenthal+mathews75}) should warn us that radiation pressure might not be negligible \citep{marconi+08}.

While the velocity-resolved reverberation mapping results were good news for the new black-hole-mass-determination industry, they created a problem for the generally accepted ``disk-wind'' explanation of the blueshiftings of high-ionization lines.  Disk-wind models are very theoretically appealing, and strong blueshiftings have been taken as signs of strong winds (e.g., \citealt{leighly+moore04}).  However, at the same time, people working on black hole mass determinations were firmly believing that they were using virialized lines!  This has almost caused AGN observers to suffer from multiple-personality disorder!\footnote{or at the very least to fear that, like the White Queen in {\it Alice in Wonderland}, they might have to believe in six impossible things before night lunch!}$^,$\footnote{The narrow-line region is creating a similar problem.  People who study extended narrow-line emission associate it with jets and outflows, while other people are using narrow-line velocity widths as a proxy for the stellar velocity dispersion (see \citealt{gaskell09a}).}

It is very difficult to finesse a disk-wind model to fit the velocity-resolved reverberation mapping constraint, all the more so since outflow was first excluded for the {\em high}-ionization C IV line \citep{gaskell88,koratkar+gaskell89,crenshaw+blackwell90,koratkar+gaskell91a, koratkar+gaskell91b, koratkar+gaskell91c}.  We believe, however, that there is a simple solution to the problem: the opacity needed to cause the blueshifting is not primarily absorption but {\em scattering} \citep{gaskell+goosmann08}.  Electron scattering had in fact been considered in the late 1960s to be a significant source of line broadening in AGNs \citep{kaneko+ohtani68,weymann70,mathis70}, but the idea fell out of favor with the success of the \citet{blumenthal+mathews75} radiative acceleration model in fitting profiles.  It has, however, long been well known (e.g., \citealt{edmonds50,auer+vanblerkom72}) that scattering off regions with a net radial motion produces line shifts.  For an infalling scattering medium, photons gain energy.  This is explained in Fig.\@ 10.  The process is similar to the well-known Fermi acceleration process.  The effect of scattering off radially moving material in AGNs was considered by \citet{kallman+krolik86} and \citet{ferrara+pietrini93}.

\begin{figure}[]
\resizebox{\hsize}{!}{\includegraphics{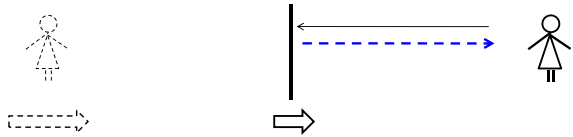}}
\caption{ \footnotesize Cartoon illustrating why scattered photons are blueshifted when scattered off a reflector which is approaching the source of photons.  The person on the right sees her reflection (far left) in the mirror.  If the mirror is approaching her, then the image is approaching her twice as quickly.} \label{}
\end{figure}

As can be seen in Fig.\@ 9, velocity-resolved reverberation mapping not only excludes outflow, but it also shows that there is a slight {\em inflow}.  Initially this result only had $\sim$90\% confidence for any one line in one object, but it has been found repeatedly for many lines in many objects now and thus, as pointed out by \citet{gaskell+snedden97}, the overall significance is high.  Early examples included \citet{gaskell88,koratkar+gaskell89,crenshaw+blackwell90,koratkar+gaskell91a, koratkar+gaskell91b, koratkar+gaskell91c,korista+95,done+krolik96,ulrich+horne96}.  More recent examples can be found in \citet{sergeev+99}, \citet{kollatschny03}, \citet{welsh+07}, \citet{doroshenko+08}, \citet{bentz+08}, \citet{denney+09} \citet{bentz+09c}.  Important independent evidence for inflow comes from high-resolution spectropolarimetry (e.g., \citealt{smith+05}).  The systematic change in polarization as a function of velocity across the Balmer lines requires a net inflow of a scattering region somewhat exterior to the Balmer lines.  Polarization reverberation mapping \citep{gaskell+07a} can reveal the location of scattering regions.

We have used the {\it STOKES} Monte Carlo radiative transfer code \citep{goosmann+gaskell07}\footnote{Available at http://www.stokes-program.info/} to model the effects on line profiles of scattering off an inflowing external medium.  The two geometries considered are shown in Fig.\@ 11.  One is an infalling spherical distribution of scatterers and the other an infalling cylindrical distribution.  In Fig.\@ 12 we show a comparison of observed profiles of two low- and high- ionization lines in PKS~0304-392 with various models with.  We adopted an infall velocity of $\sim 1000$ km s$^{-1}$ based on velocity-dependent reverberation mapping, spectropolarimetry, and the observed mean blueshift (see \citealt{gaskell+goosmann08} for details).  It can be seen that both spherically and cylindrically symmetric models readily reproduce the blueshifting.

\begin{figure}[t!]
\resizebox{\hsize}{!}{\includegraphics{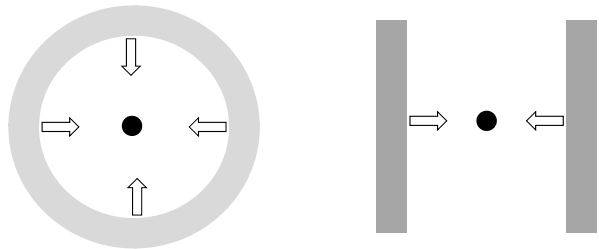}}
\caption{ \footnotesize Cross sections in a plane through the axis of symmetry of the two scattering region geometries modeled in Fig.\@ 12.} \label{}
\end{figure}

\begin{figure}[]
\resizebox{\hsize}{!}{\includegraphics{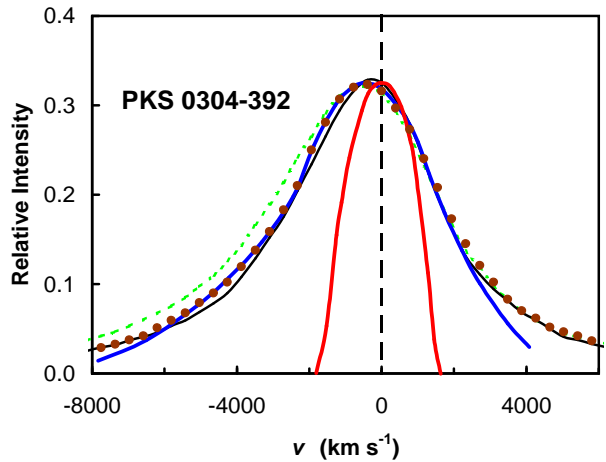}}
\caption{ \footnotesize
Modelling the blueshifting of high-ionization lines.  The profiles of O\,I $\lambda$1305 (narrow symmetric profile shown in red) and C\,IV $\lambda$1549 (thick blue line) for the quasar PKS~0304-392.  The thin black line is the blueshifted profile produced by an infalling  spherical distribution of external scatters with an electron-scattering optical depth $\tau_{es}$ = 0.5, and the dashed green line is the profile produced by the same distribution with $\tau_{es}$ = 1.  The brown dots are the profile produced by a $\tau_{es}$ = 20 infalling external cylindrical distribution.  The geometries are shown in Fig.\@ 11.  Figure from \citet{gaskell+goosmann08}; PKS~0304-392 observations taken from \citet{wilkes84}.}
\label{blueshift_example}
\end{figure}

An additional advantage of having significant scattering in the BLR is that it solves the ``smoothness problem'' for BLR line profiles \citep{capriotti+81}.  The intrinsic line broadening in an individual BLR cloudlet is only of the order of the sound speed ($\sim$ 15 km s$^{-1}$), yet the velocity broadening of the BLR as a whole is hundreds of times greater.  This requires the number of clouds to be very high \citep{capriotti+81,atwood+82}.  The limit on the number of discrete clouds has now been pushed up to $10^8$ \citep{arav+98,dietrich+99}.   This constraint is relaxed if there is broadening by scattering.

\subsection{The overall velocity field of the BLR}

For a typical AGN, several independent lines of evidence (the blueshifting, velocity-resolved reverberation mapping, and spectropolarimetry) all point to the inflow velocity being of the order of $\sim 1000$ km s$^{-1}$.  As has been mentioned, velocity-resolved reverberation mapping (see, for example, Fig.\@ 9) implies that the {\em dominant} motion is {\em not} radial, but Keplerian or random.  The observed widths of broad lines are indeed several times higher than the inflow velocity, and, of course, the predominant motion for a flattened distribution must be Keplerian.

As is clear from Fig.\@ 6 and 7, when we observe the BLR (i.e., in type-1 objects) we are always seeing it close to face-on.  The Keplerian component of velocity must be reduced by $\sin i$, where $i$ is the angle between the axis of rotational symmetry and the line of sight.  The statistics of line profiles in the SDSS \citep{lamura+09} suggest that for the vast majority of objects $i$ is $< 20\deg$).

As was realized by \citet{osterbrock78}, the statistics of line widths imply that, in addition to Keplerian motion, there has to be a substantial additional component of velocity {\em perpendicular} to the orbital plane.  Osterbrock appropriately called this ``turbulence''.
The vertical component is also necessary for the reconciling the structure of the BLR with its kinematics.  As \citet{mannucci+92} showed, for NGC~5548 the combined constraints of reverberation mapping  and  time-averaged line profile and favor the sort of ``bird's nest'' BLR distribution shown in Figs. 6 and 7.

	In summary, I believe that all the evidence points to the BLR having a nest-like appearance and having velocity components:

\begin{equation}
v_{Kepler} > v_{turb} \gtrsim v_{inflow}
\end{equation}

\noindent where the Keplerian velocity, $ v_{Kepler}$, of an emission line is a couple of times larger that the turbulent velocity, $v_{turb}$, which is in turn somewhat bigger than the inflow velocity, ${inflow}$.  The ratios of BLR height to radius and of $v_{Kepler}$ to $v_{turb}$ are similar to those deduced by \citet{osterbrock78}.  The only change to the Osterbrock model is recognizing that there is also a significant inflow.

\section{Orientation effects}

It is well established from radio properties (see \citealt{antonucci93}) that core-dominated AGNs are simply lobe-dominated AGNs viewed from near the jet axis (i.e., near face-on).  \citet{gaskell+04} showed from a comparison of continuum shapes and line ratios that core-dominated and lobe-dominated AGNs have the same underlying optical-to-UV continuum shape and that the SED differences are just due to increased reddening in the lobe-dominated AGNs.  We thus have every reason to expect the BLRs of core-dominated and lobe-dominated AGNs to be the same on average.  Lobe-dominated radio-loud AGNs should therefore be an excellent laboratory for studying how orientation affects the appearance of the BLR.  \citet{miley+miller79} found that lobe-dominated AGNs preferentially had broader and more irregular line profiles.  \citet{wills+browne86} discovered that the FWHM of H$\beta$ increases as we see AGNs more edge-on.  This provided strong support for a flattened BLR.

AGNs with the peaks of their broad Balmer lines blueshifted or redshifted from the systemic velocity have long been known \citep{lynds68}.  It was proposed \citep{gaskell83} that these peaks might represent separate BLRs each associated with a member of a supermassive black hole binary, but line profile variability observations on long and short timescales have delivered two fatal blows to this hypothesis.  Firstly, although for a while it looked like the expected binary orbital motion was being seen in long-term profile variations in 3C\,390.3 \citep{gaskell96}, further observations showed that the radial velocity changes were completely inconsistent with a binary black hole \citep{eracleous+97} but were instead consistent with orbital motion of concentrations of BLR gas orbiting in a disk.  The second fatal blow was that velocity-resolved reverberation mapping of 3C\,390.3 strongly ruled out the binary BLR hypothesis because the redshifted and blueshifted peaks varied simultaneously on a light-crossing timescale \citep{obrien+98,dietrich+98}. This demonstrated conclusively that the double-peaked profiles arose from an inclined disk, as had been widely proposed (see references and discussion in \citealt{gaskell+snedden99}). Despite these double fatal blows to the idea that displaced broad-line peaks might be due to supermassive binaries, the topic of what signs there might be of sub-parsec supermassive binaries nonetheless remains one of considerable current interest (see review by Tamara Bogdanovi\'c in these proceedings).

A subsequent comprehensive survey of radio-galaxies by \citet{eracleous+halpern94,eracleous+halpern03} revealed many disk-like Balmer line profiles.  They found the FWHMs of the Balmer lines to be  approximately double those of AGNs with single-peaked Balmer lines.  As is shown in Fig.\@ 14, a factor of two reduction in line width is sufficient to make displaced peaks disappear.  \citet{gaskell+snedden99}, \citet{popovic+04}, and \citet{bon+06} have argued that a disk-like emission line contribution is probably present in all BLRs but simply hard to recognize because, as illustrated in Fig.\@ 13, the classic double peaks become hard to see when the disk is near to face-on.

\begin{figure}
\resizebox{\hsize}{!}{\includegraphics{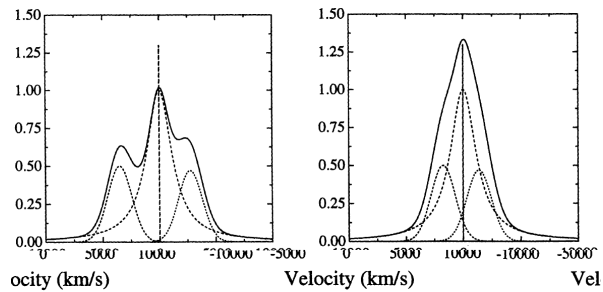}}
\caption{The effect of broadening lines on the appearance of structure in line profiles.  The left frame shows a Lorentzian and two Gaussians chosen to approximate the appearance of H$\alpha$ or H$\beta$ in 3C\,390.3 in 1981 or 1988.  The right frame has the same line widths and peak intensities as in the left frame, but half the velocity displacements.  Figure from \citet{gaskell+snedden97}.}
\label{}
\end{figure}

It is straight forward to estimate the inclinations of the BLRs from broad disk-like line profiles.  These can often be estimated to within a few degrees.  \citet{eracleous+halpern94,eracleous+halpern03} get inclinations which predominantly have $i > 25\deg$.  Their fits to the disk profiles also provide important confirmation that a significant turbulent velocity is needed and give the turbulent velocity for each object.  Without the turbulent velocity component the peaks of the line profiles would be much too sharp.  The turbulent velocities are fairly well determined from the line profile fits (to $\approx \pm 250$ km s$^{-1}$).  The average BLR turbulent velocity needed is 1300 km s$^{-1}$.  This is roughly what would be expected from the height of the BLR/torus.  The 1-$\sigma$ scatter in the derived turbulent velocities is only $\pm 400$ km s$^{-1}$, which is only slightly greater than the average formal uncertainty in the estimates.

\citet{bon08} has estimated inclinations for single-peaked AGNs.  For these we mostly see disks with inclinations of $i < 25\deg$ (see also Bon et al. in these proceedings).  The difference in $\sin i$ between the displaced-BLR-peak AGNs and single-peaked AGNs is thus about a factor of two.  This agrees with the ratios of FWHMs for the two samples.

\section{The accuracy of AGN black hole mass determinations}

The component of velocity perpendicular to the equatorial plane is {\em vital} for AGN black hole mass determinations!  Without this strong turbulent velocity component, variations in $\sin i$ would introduce {\em substantial} scatter into AGN black hole mass estimates, especially since type-1 AGNs are observed close to face-on.  There is recent  evidence that there is remarkably little scatter in AGN mass estimates.  Firstly, it has become apparent \citep{bochkarev+gaskell09} that the two main methods of estimating black hole masses from the BLR agree surprisingly well. The Dibai single-epoch-spectrum method \citep{dibai77} and reverberation mapping methods agree to within the expected errors. \citet{gaskell09b} has shown furthermore that a simple refinement of the method produces even better agreement.  The agreements mean that such methods are estimating the effective radii of the BLR correctly.  As \citet{bochkarev+gaskell09} discuss, the success of the Dibai method means that the inner regions of AGNs are very similar.   In particular:

\begin{enumerate}

\item The spectral energy distribution (SED) from the optical to the far UV must be very similar in all type-1 AGNs because the optical region where the flux is measured is far removed in energy from the far UV which is photoionizing the gas.   Although AGN SEDs {\em look} different, \citet{gaskell+04} and \citet{gaskell+benker07} have already argued that the apparent variation is not real but is primarily caused by reddening.

\item There is a simple scaling relationship between the luminosity and the effective radius. This is supported by reverberation mapping estimates of the effective radii of BLRs \citep{koratkar+gaskell91c,kaspi+00,kaspi+05,bentz+06,bentz+09a}.

\end{enumerate}

Both the Dibai and reverberation-mapping methods of estimating black hole masses depend on observed BLR line widths, so geometric differences and orientation effects will affect {\em both} methods.  An important {\em  external} check on the accuracy of AGN black hole mass estimates is provided by the tightness of the relationship between black hole mass, $M_\bullet$, and luminosity, $L_{host}$, of the bulge of the host galaxy.  \citet{gaskell+kormendy09} have recently shown that estimating $M_\bullet$ by the Dibai method and $L_{host}$ from the fraction of starlight in SDSS spectra gives a scatter of $\pm 0.23$ dex in $\log M_\bullet$ (see Fig.\@ 14).   \citet{bentz+09b} have estimated $L_{host}$ completely independently for a different set of AGNs using HST photometry and published reverberation mapping mass estimates.  They get a scatter in $\log M_\bullet$ of $\pm 0.33$ dex.  Both of these scatters in the AGN $ M_\bullet$ -- log$L_{host}$ relationships are smaller than the $\pm 0.38$ dex scatter \citet{gultekin+09} and others find when $M_\bullet$ is determined by stellar dynamical methods, but they are still greater than the $\pm 0.17$ dex scatter in the $M_\bullet$\,--\,$\sigma_*$ relationship for pure bulge (i.e., barless) galaxies \citep{graham08}.  The Dibai method and the method proposed by \citet{gaskell09b} seems to give particularly tight $M_\bullet$\,--\,$\sigma_*$ and $M_\bullet$\,--\,$L_{bulge}$ relationships for the most massive elliptical galaxies \citep{gaskell09a}.  This is probably because they have the least intrinsic scatter in the $M_\bullet$\,--\,$\sigma_*$ relationship. These comparisons with predictions from host galaxy properties imply that black hole mass determinations from the BLR are surprisingly accurate -- as accurate as the best stellar-dynamical estimates.  This accuracy of black hole mass estimates made using the BLR provides strong support for all type-1 AGNs being very similar as far as the structure and kinematics of the BLR goes, and for orientation effects being minimal.  The accuracy of AGN black hole mass estimates is thus consistent with there being a substantial turbulent BLR velocity component and type-1 AGNs being seen close to pole-on.

\begin{figure}
\resizebox{\hsize}{!}{\includegraphics{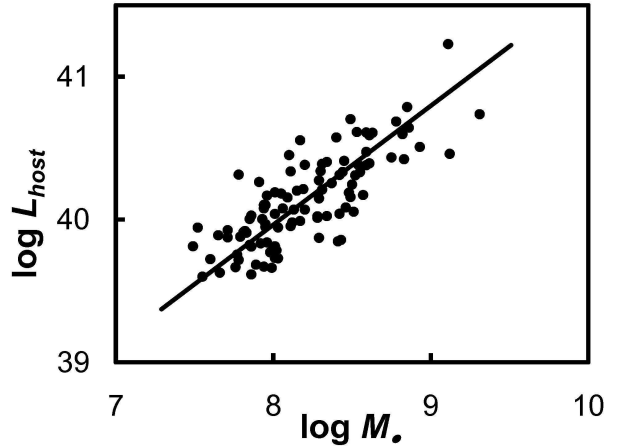}}
\caption{The $M_\bullet$ -- $L_{host}$ relationship for 100 AGNs with $0.13 < z < 0.34$. See \citet{gaskell+kormendy09} for details.   $M_\bullet$ has been estimated by the Dibai method.  The diagonal line is the OLS-bisector fit, $M_\bullet \propto L^{0.84}_{host}$.}
\label{}
\end{figure}

\section{What drives BLR motions?}

The circular component of motion of BLR clouds is a simple consequence of gravity, but the turbulent vertical component of motion has to be maintained against dissipation losses, and an outward transfer of angular momentum is necessary to get inflow.  We now know that the viscosity needed to drive the outward flow of angular momentum in accretion discs, and hence the inward flow of matter, is the magneto-rotation instability (MRI)  \citep{balbus+hawley91}.  Over the last decade increases in computing power and the development of more sophisticated programs by several groups have allowed increasingly detailed magneto-hydrodynamic (MHD) simulations of accretion flows (e.g., \citealt{hawley+krolik01,proga03,anninos+05,ohsuga+09,shafee+08}, and references therein).  In these models attention has been focused a lot on the low-density outflows.  Because emissivity goes as the square of density, the emission is dominated by the high-density.  To my mind what is impressive about every single one of these models is that, despite the different modeling approaches, {\em the velocity fields of the high-density material all match the velocity field inferred for the BLR}.  I.e., the dominant motion is Keplerian, but there is substantial turbulence, and a significant inflow.  This is very clear when one watches movies different groups make of their simulations.   I believe that these simulations give us a physical basis for what we have deduced from BLR observations: the BLR {\em is} the material accreting onto the black holes.  It was indeed noted a long time ago that if the BLR is inflowing it can provide the necessary mass flux for powering the AGN \citep{padovani+rafanelli88}.

\section{Conclusions and unsolved problems}

I think we now have a fairly good emerging picture of what the BLR is like and what role it plays in the life of an AGN.  Interstellar material approaching the nucleus settles into a flattened distribution, the thick torus.  Material loses angular momentum because of MRI turbulence and gradually spirals inwards.  When the material of the torus gets within the dust sublimation radius, the dust evaporates and we have the BLR.  The turbulent BLR continues to spiral inwards towards the black hole where it is eventually accreted.  The degree of ionization increases as the gas gets closer in.  The optically-thick material, which will tend to be concentrated towards the mid-plane, produces continuum emission; the more optically-thin material produces the BLR.   Not all of the BLR is accreted.  Some of it is driven off the surface of the BLR/torus in a high-velocity, low-density wind, as is found in all the MHD simulations and as is observed.

Although I believe we are getting a clear overall picture of the BLR, there is still plenty to work on both observationally and theoretically!  For a start, the picture discussed above needs to be thoroughly tested to verify that it works for all objects and not just a few well-observed objects such as NGC\,5548.  More work needs to be done to see whether a disk-wind model (the leading rival to the model presented here) could also explain everything.  We know that there is outflowing gas as well as gas accreting onto the black hole.  The question is: how much of this is also contributing to the broad-line profiles? (especially to the high-ionization lines).  \citet{ilic+08}, for example, have shown that outflows can match some observed BLR profiles, so determining the relative contribution of an outflow to broad-line profiles from line-profile fitting alone is difficult.  I think that reverberation mapping and spectropolarimetry (see, for example, \citealt{axon+08}) are going to provide the best answers.  \citet{kollatschny03} found marginal evidence for some outflow in Mrk 110, and \citet{denney+09b} have found a clearer signature of an apparent outflow component in velocity-resolved reverberation mapping of NGC\,3227.  Since the \citet{denney+09b} results are from a single short observing campaign, I do not think that NGC\,3227 presents a major problem yet for the general picture presented here.\footnote{The uncertainties in the red-wind/blue-wing lags can be larger than thought.  NGC\,5548 provides a good illustration of this. The red-wing/blue-wing lag varies from year to year by more than the formal errors \citep{welsh+07}, but the NGC\,5548 BLR is probably not changing direction at the end of every observing season!  A strong reason for believing that the NGC\,3227 kinematics are {\em not} unusual is that, as \citet{denney+09b} point out, the mass estimate lies on the $M_\bullet$\,--\,$\sigma_*$ relationship.} If follow-up observing campaigns confirm the signature of outflow in NGC\,3227 then this would be a significant challenge to the model favored here. Nevertheless, the \citet{denney+09b} result does caution us that AGNs might not all be identical in the relative dominance of inflow and outflow.

Even if we are right about the basic structure of the BLR and torus of AGNs, there are still a lot of interesting and potentially important details in need of further investigation.  Although in this review I have been emphasizing the similarities among AGNs and what they imply, there are some significant {\em differences} in the BLRs too (see, for example, \citealt{marziani+95}).  If our basic framework of how an AGN works is correct, then the differences need to be explicable within the framework too.  Space here only permits a brief mention of some of these problems, but fortunately many of them are reviewed and discussed elsewhere in these proceedings (see, for example, the reviews by Mike Eracleous and Jack Sulentic).

It has been known for over three decades now that object-to-object differences are correlated with each other, and one of main drivers of the correlated differences is the Eddington ratio (see \citealt{sulentic+00} and these proceedings).  Since we now have reliable AGN black holes masses, we also have reliable Eddington ratios, so there is a lot that can be done in investigating the dependence of BLR properties on accretion rate.  I think there is a lot that needs explaining here.

The biggest object-to-object difference in optical spectra is optical Fe\,II emission \citep{osterbrock77}.  Understanding how the very strong optical Fe\,II emission seen in AGNs is produced has been a long-standing problem (see \citealt{baldwin+04,joly+08,hu+08,kuehn+08,verner+09,dong+09} for recent discussions).  In the BLR model discussed here, optical Fe\,II emission arises in the outer part of the BLR just inside the torus (and quite likely overlapping with it), but this does not readily explain why optical Fe\,II is so much stronger in some objects than others.

Another mystery of the correlated object-to-object differences is the strength of the {\em narrow}-line region (NLR) emission.  This is the other strong object-to-object difference and it is mysteriously strongly anti-correlated with Fe\,II emission \citep{osterbrock77,steiner81,boroson+oke84,gaskell87,boroson+green92}.  A complete model of AGNs needs to explain why the NLR and BLR know about each other.

Although I have argued that the basic properties of AGNs with broad disk-like Balmer line profiles are consistent with the picture presented here, these objects, and especially the variability of their profiles, present some special challenges, as is discussed in Mike Eracleous's review.  There is also a lot more to be learned with orientation effects.

In summary, I think that although our overall picture of the BLR and the role it plays in the AGN phenomenon is becoming clearer, many mysteries remain, there is still a lot to learn, and there are probably surprises in store.

\vspace{0.7cm}

\noindent I am grateful to Luka Popovi\'c, Milan Dimitrijevi\'c, and the other members of the scientific organizing committee of the 7th Serbian Conference on Spectral Line Shapes for inviting me to speak on this topic.  I would like to express my appreciation to Dragana Ili\'c and all the members of the local organizing committee for providing a very pleasant and stimulating experience throughout the conference, both culturally and scientifically.  I also have to thank my collaborators and former graduate students for all their contributions and discussions over the years, and the anonymous referee for useful comments.  This research has been supported in part by US National Science Foundation grant AST 08-03883.

\end{document}